\documentclass[12pt]{article}
\usepackage{graphicx}
\date{}
\def\be{\begin{equation}}
\def\ee{\end{equation}}
\def\bea{\begin{eqnarray}}
\def\eea{\end{eqnarray}}
\def\s{\sigma}
\def\al{\alpha}

\def\de{\delta}
\def\om{\omega}
\def\pr{\prime}
\def\f{\varphi}
\def\ep{\varepsilon}
\def\tom{\tilde\omega}

\title{CLOSED STRING WITH MASSES  \\
IN MODELS OF BARYONS AND GLUEBALLS }
\author{G.\,S. Sharov\\
{\small Tver state university}\\
{\small Tver, 170002, Sadovyj per. 35, Mathem. dep-t}}
\begin{document}
\maketitle
\begin{abstract}
The closed string carrying $n$ point-like masses is considered as
the model of a baryon ($n=3$), a glueball ($n=2$ or 3)  or another
exotic hadron. For this system the rotational states are obtained
and classified. They correspond to exact solutions of dynamical
equations, describing an uniform rotation of the string with
massive points. These rotational states result in a set of
quasilinear Regge trajectories with different behavior.

The stability problem for the so called central rotational states
(with a mass at the rotational center) is solved with using the
analysis of small disturbances. These states turned out to be
unstable, if the central mass is less than some critical value.
\end{abstract}

\centerline {\bf 1. Introduction}
\medskip

The Nambu-Goto string (or relativistic string) simulates strong
interaction between quarks at large distances in various string
models of mesons and baryons \cite{Nambu}\,--\,\cite{Solovm}
(Fig.\,1{\it a\,--\,e}). This string has linearly growing energy
(energy density is equal to the string tension $\gamma$) and
describes nonperturbative contribution of the gluon field and the
QCD confinement mechanism.

Such a string with massive ends in Fig.\,1{\it a} may be regarded
as the meson string model \cite{Ch,BN}. String models of the
baryon were suggested in the following four topologically
different variants \cite{AY}: ({\it b}) the quark-diquark model
$q$-$qq$ \cite{Ko} (on the classic level it coincides with the
meson model ({\it a}), ({\it c}) the linear configuration
$q$-$q$-$q$ \cite{lin}, ({\it d}) the ``three-string'' model or Y
configuration \cite{AY,PY}, and (e) the ``triangle'' model or
$\Delta$ configuration \cite{Tr,PRTr}.

All cited string hadron models generate linear or quasilinear Regge
trajectories in the limit of large energies for excited states of
mesons and baryons \cite{4B,Ko,InSh,Solovm}
 \be
 J\simeq\al_0+\al'E^2,
 \label{Reggm} \ee
 if we use rotational states of these systems (planar uniform
rotations). Here $J$ and $E$ are the angular momentum and energy
of a hadron state (or rotational state of a string model), the
slope $\al'\simeq0{.}9$ GeV$^{-2}$. For the model of meson in
Fig.\,1{\it a} and for the quark-diquark baryon model in
Fig.\,1{\it b} this slope and the string tension $\gamma$ are
connected by the Nambu relation \cite{Nambu}
 \be
 \alpha'=\frac1{2\pi\gamma}.
 \label{Namb}\ee
 The same relation (\ref{Namb}) takes place for the
linear baryon configuration (Fig.\,1{\it c}) in the case of its
central rotational states. In these states the middle mass is at
the rotational center. In papers \cite{lin,stab} we have shown
that the mentioned states are unstable with respect to small
disturbances.

\begin{figure}[ht]
\includegraphics[scale=0.85,trim=60 30 10 170]{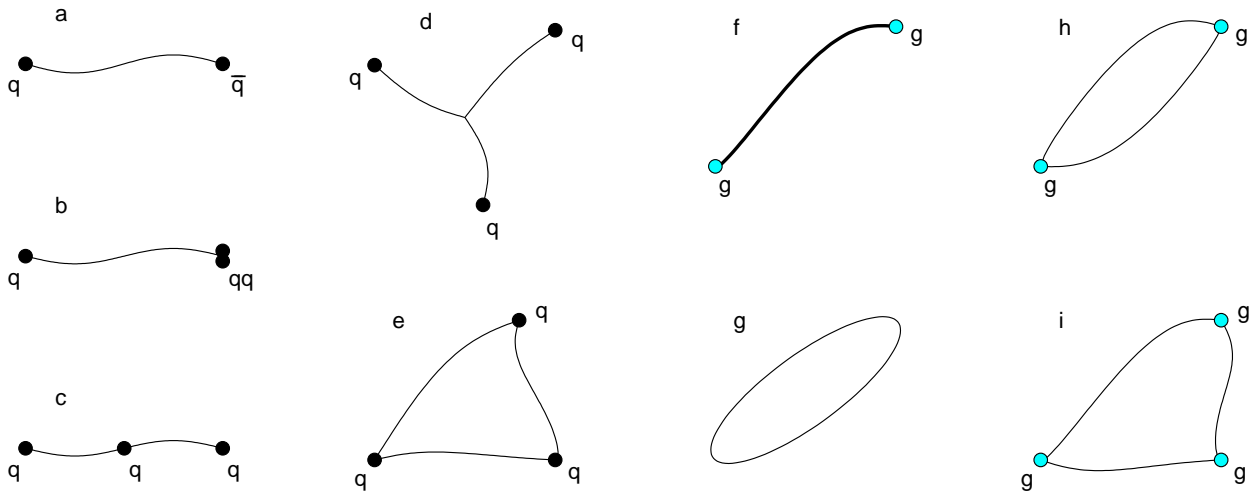}
\caption{String models of mesons, baryons and glueballs}
\end{figure}

The string baryon model Y (Fig.\,1{\it d}) for its rotational
states demonstrates the Regge asymptotics (\ref{Reggm}) with the
slope \cite{PY} $\alpha'=1/(3\pi\gamma)$.
 To describe the experimental Regge trajectories with the slope
$\al'\simeq0{.}9$ GeV$^{-2}$ we are to assume that the effective
string tension $\gamma_Y$ in this model differs from the value
$\gamma$ in models in Figs.\,1{\it a\,--\,c} (the fundamental
string tension) \cite{4B,InSh} and equals
$\gamma_Y=\frac23\gamma.$
 Moreover, the rotations of the Y string configuration are also unstable
with respect to small disturbances on the classic level
\cite{stab,Y02}.

 In this paper we'll concentrate on applications of the
closed string carrying $n$ point-like masses in hadron
spectroscopy. The string baryon model ``triangle'' or $\Delta$ is
the particular case of this system if $n=3$. This model generates
a set of rotational states with different topology \cite{Tr,PRTr}.
The so called triangle states was applied for describing excited
baryon states on the Regge trajectories \cite{4B,InSh}, but in
this case (like for the model Y) we are to take another effective
string tension $\gamma_\Delta=\frac38\gamma$.

Different string models were used for describing glueballs (bound
states of gluons) \cite{Solovgl}\,--\,\cite{MathSS3} and other
exotic hadrons \cite{KrMart}. There are a lot of experimentally
observed hadron states which may be interpreted as glueballs
\cite{Anis,KlemptZ} predicted in QCD. But these states can mix
with meson states, so their glueball interpretation is ambiguous
one.

String models of glueballs or some exotic hadron states (glueball
candidates) were suggested in the following variants
\cite{Solovgl}\,--\,\cite{MathSS3} shown in Fig.\,1: ({\it f}) the
open string with enhanced tension (the adjoint string) and two
constintuent gluons at the endpoints; ({\it g}) the closed string
simulating gluonic field; ({\it h}) the closed string carrying two
point-like masses (constituent gluons). Evidently, the last model
may be easily generalized for three-gluon glueballs \cite{MathSS3}
in Fig.\,1{\it i}. The cited authors suggested different
approaches. Some of them \cite{AbreuB}\,--\,\cite{MathSS3} used
potential models with string term in the potential.

Glueballs, their masses and momenta, corresponding Regge
trajectories are simulated in lattice calculations
\cite{MorningP}\,--\,\cite{Meyerth}. In this approach the QCD
glueball may be identified with the pomeron that is the Regge pole
determining an asymptotic behavior of high-energy diffractive
amplitudes \cite{Meyerth}\,--\,\cite{Kaidal01}. The pomeron Regge
trajectory \cite{MeyerT,Meyerth,DonPom}
 \be
 J\simeq1{.}08+0{.}25E^2
 \label{ReggPom} \ee
 differs from hadronic ones (\ref{Reggm}):
its slope $\alpha'\simeq0{.}25$ is essentially lower, and its
intercept $\alpha_0\simeq1{.}08$ is positive and rather large.

The mentioned string models of glueballs in in Fig.\,1{\it
f\,--\,i} were suggested, in particular, to describe Regge
trajectories of the type (\ref{ReggPom}).

The model in Fig.\,1{\it f} considered in
Refs.~\cite{EstrCBRS,SzczS,Meyerth,Sim90}, is the open string with
massive ends (describing gluons). This string (the adjoint string)
has the following tension \cite{Meyerth,Bali0}
 $$\gamma_{adj}=\frac{2N_c^2}{N_c^2-1}\gamma=\frac94\gamma.
 $$
 It exceeds tension $\gamma$ in hadron models in Fig.\,1{\it a\,--\,c}
(fundamental string). In accordance with the Nambu relation
(\ref{Namb}) the corresponding slope of Regge trajectories is
$\frac49$ of this value for mesons: $\alpha'\simeq0{.}4$. It is
larger than the slope $\alpha'\simeq0{.}25$ of the trajectory
(\ref{ReggPom}).

In papers \cite{Solovgl,ZayasS,PonsRT,MeyerT,Meyerth} the closed
string without masses (Fig.\,1{\it g}) is used for describing
glueballs. The authors chose different classes of classic motions
(states) for modelling trajectories (\ref{ReggPom}) for glueballs.
In Ref.~\cite{Solovgl} this class includes rotations and
oscillations of the elliptic closed Nambu-Goto string. This
results to a set of slopes of Regge trajectories from 0 to
$(4\pi\gamma)^{-1}$. In Refs.~\cite{MeyerT,Meyerth} the circular
shape of the string is fixed and the spectrum of its excitations
is considered. In Refs.~\cite{ZayasS,PonsRT,MilSh} the closed
string is embedded into spaces with nontrivial geometry.

The closed string carrying two massive points (Fig.\,1{\it h})
(they describe constituent gluons) in Ref.~\cite{AbreuB} is taken
as a basis of the potential model of a glueball. The similar
potential model of the 3-gluon glueball is constructed in
Ref.~\cite{MathSS3}. It corresponds to the system in Fig.\,1{\it
i}. The latter model is similar to the string baryon
model``triangle'' \cite{Tr,PRTr} in Fig.\,1{\it e}.
\cite{Tr,PRTr}. But the configurations in Figs.\,1{\it h} and {\it
i} are not investigated yet as string models of a glueball.

In this paper we consider classical dynamics of the closed string
carrying $n$ massive points (generalization of the models in
Figs.\,1{\it e, h} and {\it i}) in Minkowski space $R^{1,3}$. In
Sect.~3 rotational states (planar uniform rotations) of this
system are described and classified. They have much more
complicated structure than a well known set of rotations of the
folded rectilinear string. All rotational states are divided into
3 classes: linear, hypocycloidal and central states (in the last
case a mass is at the rotational center).  In Sect.~4 the
stability problem for central rotational states is solved with
using the analysis of small disturbances.
 Rotational states of string systems are
widely used for generating Regge trajectories. Their structure and
behavior for the considered system are described in Sect.~5.

\bigskip

\centerline {\bf 2. Dynamics}
\medskip

The dynamics of the closed string carrying $n$ point-like masses
$m_1$, $m_2,\;\dots\,m_n$ is determined by the action
 \be
S=-\gamma\int\limits_{\Omega}\sqrt{-g}\;d\tau d\s
-\sum\limits_{i=1}^n m_i\int\sqrt{\dot x_i^2(\tau)}\;d\tau,
 \label{S}\ee
 generalizing the case of the string baryon model ``triangle''
 \cite{PRTr}. Here $\gamma$ is the string tension, $g$ is the determinant of the
induced metric
$g_{ab}=\eta_{\mu\nu}\partial_aX^\mu\partial_bX^\nu$ on the string
world surface $X^\mu(\tau,\s)$, the speed of light $c=1$. The
world surface mapping in  $R^{1,3}$ from
$\Omega=\{\tau,\s:\,\tau_1<\tau<\tau_2,\,\s_0(\tau)<\s<\s_n(\tau)\}$
is divided by the world lines of massive points
$x_i^\mu(\tau)=X^\mu(\tau,\s_i(\tau))$, $i=0,\dots,n$ into $n$
world sheets. Two of these functions $x_0(\tau)$ and $x_n(\tau)$
describe the same trajectory of the $n$-th massive point, and
their equality forms the closure condition
 \be
X^\mu(\tau,\s_0(\tau))=X^\mu(\tau^*,\s_n(\tau^*))
 \label{clos}\ee
 on the tube-like world surface \cite{PRTr,MilSh}. These equations may
contain two different parameters $\tau$ and $\tau^*$, connected
via the relation $\tau^*=\tau^*(\tau)$. This relation should be
included in the closure condition (\ref{clos}) of the world
surface.

Equations of motion of this system result from the action (\ref{S})
and its variation. They may be reduced to the simplest form under
the orthonormality conditions on the world surface
 \be
(\partial_\tau X\pm\partial_\s X)^2=0,
 \label{ort}\ee
 and the conditions
 \be \s_0(\tau)=0,\qquad \s_n(\tau)=2\pi.
\label{ends}\ee
 Conditions (\ref{ort}), (\ref{ends}) always may be
fixed without loss of generality, if we choose the relevant
coordinates $\tau$, $\s$ \cite{PRTr}. It is connected with the
invariance of the action (\ref{S}) with respect to nondegenerate
reparametrizations on the world surface
$\tau=\tau(\tilde\tau,\tilde\s)$, $\s=\s(\tilde\tau,\tilde\s)$.
The scalar square in Eq.~(\ref{ort}) results from scalar product
$(\xi,\zeta)=\eta_{\mu\nu}\xi^\mu\zeta^\nu$.

The orthonormality conditions (\ref{ort}) are equivalent to the
conformal flatness of the induced metric $g_{ab}$. Under conditions
(\ref{ort}), (\ref{ends}) the string motion equations take the form
 \cite{4B,PRTr,MilSh}
 \be \frac{\partial^2X^\mu}{\partial\tau^2}-
\frac{\partial^2X^\mu}{\partial\s^2}=0,
 \label{eq}\ee
 \be m_i\frac d{d\tau}\frac{\dot
x_i^\mu(\tau)}{\sqrt{\dot x_i^2(\tau)}}+\gamma
\Big[X^{'\!\mu}+\dot\s_i(\tau)\dot
X^\mu\Big]\Big|_{\s=\s_i-0}-\gamma\Big[X^{'\!\mu}+\dot\s_i(\tau)\dot
X^\mu\Big]\Big|_{\s=\s_i+0}=0,
 \label{qqi}\ee
 \be  m_n\frac d{d\tau}\frac{\dot x_0^\mu(\tau)}{\sqrt{\dot
x_0^2(\tau)}}+\gamma
\big[X^{'\!\mu}(\tau^*,\sigma_n)-X^{'\!\mu}(\tau,0)\big]=0.
\label{qq0} \ee
 Here $i=1,\dots,n-1$, $\dot X^\mu\equiv\partial_\tau X^\mu$,
$X^{'\!\mu}\equiv\partial_\s X^\mu$.

Eqs.~(\ref{qqi}), (\ref{qq0}) are equations of motion for the
massive points resulting from the action (\ref{S}). They may be
interpreted as boundary conditions for Eq.~(\ref{eq}).

We denote the unit vectors
 $$ e^\mu_0,\;e^\mu_1,\;e^\mu_2,\;e^\mu_3,$$
 associated with coordinates $x^\mu$.
These vectors form the orthonormal basis in $R^{1,3}$.

The system of equations (\ref{clos})\,--\,(\ref{qq0}) describe
dynamics of the closed string carrying $n$ point-like masses
without loss of generality. One also should add that the tube-like
world surface of the closed string is continuous one, but its
derivatives may have discontinuities at the world lines of the
massive points (except for derivatives along these lines)
\cite{PRTr}. These discontinuities are taken into account in
Eqs.~(\ref{qqi}), (\ref{qq0}).

\bigskip

\centerline {\bf 3. Rotational states}
\medskip

We search rotational solutions of system
(\ref{clos})\,--\,(\ref{qq0}) using the approach supposed in
Ref.~\cite{PRTr} for the string model ``triangle'' and in
Ref.~\cite{MilSh} for the closed string carrying one massive
point. In the frameworks of the orthonormality gauge (\ref{ort})
we suppose that the system uniformly rotates, masses move at
constant speeds $v_i$ along circles and conditions
 \be
\s_i(\tau)=\s_i={}\mbox{const},\qquad  i=1,\dots,n,
 \label{sigmi}\ee
 \be
 \tau^*=\tau+2\pi\theta,\qquad\theta={}\mbox{const},
 \label{theta}\ee
 \be
\frac\gamma{m_i}\sqrt{\dot X^2(\tau,\s_i)}=Q_i={}\mbox{const},
\qquad i=1,\dots,n
 \label{Qi}\ee
 are fulfilled.

When we search solution of the linearized system
(\ref{clos})\,--\,(\ref{qq0}) under restrictions
(\ref{sigmi})\,--\,(\ref{Qi}) as a linear combination of terms
$X^\mu(\tau,\s)= T^\mu(\tau)\,u(\s)$ (Fourier method) we obtain
from Eq.~(\ref{eq}) two equations for functions $T^\mu(\tau)$ and
$u(\s)$:
 $$
T_\mu^{\pr\pr}(\tau)+\om^2 T_\mu=0,\qquad  u''(\s)+\om^2 u=0.$$

Their solutions describing uniform rotations of the string system
(rotational states) contain one nonzero frequency $\om$ and have the
following form \cite{PRTr,MilSh}:
 \be
X^\mu(\tau,\s)=x_0^\mu+e_0^\mu(a_0\tau+b_0\s)+ u(\s)\cdot
e^\mu(\om\tau)+\tilde u(\s)\cdot\acute e^\mu(\om\tau).
 \label{X}\ee

Here
 $$e^\mu(\om\tau)=e^\mu_1\cos\om\tau+e^\mu_2\sin\om\tau,\qquad
\acute e^\mu(\om\tau)=-e^\mu_1\sin\om\tau+e^\mu_2\cos\om\tau$$
 are unit orthogonal vectors rotating in the plane $e_1,\,e_2$;
the function
 $$
u(\sigma)=\left\{\begin{array}{ll}A_1\cos\omega\sigma+B_1\sin\omega\sigma,&
\sigma\in[0,\sigma_1],\\
A_2\cos\omega\sigma+B_2\sin\omega\sigma,&\sigma\in[\sigma_1,\sigma_2],\\
\dots & \;\\
A_n\cos\omega\sigma+B_n\sin\omega\sigma,&\sigma\in[\sigma_{n-1},2\pi]
\end{array}\right.
 $$
 and its analog
 $$\tilde u(\sigma)=
\tilde A_i\cos\omega\sigma+\tilde
B_i\sin\omega\sigma,\quad\sigma\in[\sigma_{i-1},\sigma_i]$$
 are continuous, but their derivatives
 have discontinuities at $\s=\s_i$ (positions of masses $m_i$).

Continuity of functions $u(\s)$ and $\tilde u(\s)$ at $\s=\s_i$
results in equalities
 \be
({\cal A}_{i+1}-{\cal A}_i)\cos\omega\s_i= ({\cal B}_i - {\cal
B}_{i+1})\sin\omega\s_i.
 \label{ABcont}\ee
 Here we use the notations for columns
 \be
{\cal A}_i=\left(\begin{array}{c} A_i\\ \tilde
A_i\end{array}\right),\qquad {\cal B}_i=\left(\begin{array}{c} B_i\\
\tilde B_i\end{array}\right).
 \label{ABmat}\ee

Expression (\ref{X}) is the solution of Eq.\,(\ref{eq}) and it
must satisfy the conditions (\ref{clos}), (\ref{ort}),
(\ref{qqi}), (\ref{qq0})
 under restrictions (\ref{sigmi})\,--\,(\ref{Qi}).
Boundary conditions (\ref{qqi}) with adding Eq.~(\ref{Qi}) take
the form
 $$
\ddot
X(\tau,\s_i)+Q_i\Big[X^{'\!\mu}(\tau,\s_i-0)-X^{'\!\mu}(\tau,\s_i+0)\Big]=0,\quad
i=1,\dots,n-1.
 $$
 Substituting Eq.\,(\ref{X}) into this relation we obtain the
equations for the columns (\ref{ABmat})
 \be
({\cal A}_{i+1}-{\cal A}_i-h_i{\cal B}_i)\,\breve S_i= ({\cal
B}_{i+1}-{\cal B}_i+h_i{\cal A}_i)\,\breve C_i.
 \label{A2}\ee
 Here and below we denote the constants
 \be
h_i=\frac\om{Q_i}=\frac{\om m_i}\gamma\Big[\dot
X^2(\tau,\s_i)\Big]^{-1/2},
 \label{hi}\ee
 $$\breve C_i=\cos\omega\sigma_i,\quad\breve
 S_i=\sin\omega\sigma_i,\;\quad
C_i =\cos\omega(\sigma_i-\sigma_{i-1}),\quad S_i
=\sin\omega(\sigma_i-\sigma_{i-1}).$$

 One can express  the columns (\ref{ABmat})  ${\cal A}_{i+1}$, ${\cal
B}_{i+1}$ of functions $u$ and $\tilde u$ in the segment
$[\sigma_i,\sigma_{i+1}]$ (between masses $m_i$ and $m_{i+1}$) via
the similar coefficients in the $[\sigma_{i-1},\sigma_i]$:
 \be
 \begin{array}{c}
 {\cal A}_{i+1}= (1+h_i\breve C_i\breve S_i)\,{\cal A}_i+h_i\breve S_i^2{\cal B}_i,\\
 {\cal B}_{i+1}= -h_i\breve C_i^2{\cal A}_i+(1-h_i\breve C_i\breve S_i)\,{\cal B}_i.
  \rule{0mm}{1.2em}
 \end{array}\label{AiBi}\ee
 Hence, al mentioned coefficients may be expressed via
${\cal A}_1$, ${\cal B}_1$.

Substituting expression (\ref{X}) into the closure condition
(\ref{clos}) and into the $n$-th boundary condition (\ref{qq0})
and keeping in mind Eqs.~(\ref{theta}), (\ref{Qi}), we obtain the
following relations for amplitudes:
 \be
b_0=-\theta a_0,
 \label{ba0}\ee
  \be
\begin{array}{c}
{\cal A}_1=M_\theta(\breve C{\cal A}_n+\breve S{\cal B}_n),\\
 {\cal B}_1=M_\theta\big[(\breve C-h_n\breve S)\,{\cal B}_n
 -(\breve S+h_n\breve C)\,{\cal A}_n\big].
 \rule{0mm}{1.2em}
 \end{array} \label{ABn}\ee
 Here $\;\breve C\equiv\breve C_n=\cos2\pi\omega$, $\;\breve S\equiv\breve S_n=\sin2\pi\omega$,
$$\!
M_\theta=\left(\begin{array}{cc} C_\theta & -S_\theta\\ S_\theta &
C_\theta,\end{array}\right), \;\;
C_\theta=\cos2\pi\theta\omega,\;\; S_\theta=\sin2\pi\theta\omega.
$$

The system of matrix equations (\ref{AiBi}), (\ref{ABn}) is
homogeneous one. It can be reduced after excluding factors ${\cal
A}_i$, ${\cal B}_i$ with $i=2,\;3,\dots,n$ to the form
 \be M_1{\cal A}_1=M_2{\cal B}_1,\qquad M_3{\cal
A}_1=M_4{\cal B}_1,
 \label{M14}\ee
 where matrices $M_k$ are linear combinations of $M_\theta$ and
the identity matrix $I$. In particular, for $n=2$ they are
 $$\begin{array}{ll}
M_1=(\breve C-h_1C_1S_2)\,M_\theta-I,&
M_2=-(\breve S-h_1S_1S_2)\,M_\theta,\\
M_3=(\breve S+h_1C_1C_2)\,M_\theta+h_2I,&
 M_4=(\breve C-h_1S_1C_2)\,M_\theta-I.
\end{array}$$

Taking into account mutual commutativity of the matrices $M_k$ and
excluding the column ${\cal B}_1$ (or ${\cal A}_1$) from the
system (\ref{M14}) we obtain the system equivalent to
Eqs.~(\ref{M14})
 \be
M{\cal A}_1=0,\qquad M{\cal B}_1=0.
 \label{M0}\ee
 Here the matrix $M=M_1M_4-M_2M_3$ may be reduced with using
equality $M_\theta^2=2C_\theta M_\theta-I$. The system (\ref{M0})
(or (\ref{M14})) has nontrivial solutions if and only if $\det
M=0$ or
 \be
 2(C_\theta-\breve C)+\breve S\sum_{i=1}^nh_i-
\sum_{i<j}h_ih_js_{ji}s_{ij}+\sum_{i<j<k}h_ih_jh_k
s_{ji}s_{kj}s_{ik}-\dots+(-1)^{n+1}\prod_{i=1}^nh_iS_i=0,
 \label{eqCg} \ee
 where $s_{ji}=\left\{ \begin{array}{ll}\sin\omega(\sigma_j-\sigma_i),& j>i,\\
 \sin\omega(2\pi+\sigma_j-\sigma_i),& j<i, \end{array}\right.$.
 Eq.~(\ref{eqCg})
connects unknown (for the present moment) values of parameters
$\om$, $\theta$, $\s_i$, $Q_i$, $h_i$.

In the case $n=2$ equation (\ref{eqCg}) is
 \be
2(C_\theta-\breve C)+(h_1+h_2)\,\breve S-h_1h_2S_1S_2=0,
 \label{eqC}\ee
This equation may be rewritten after expanding notations
 $$ 2(\cos2\pi\theta\omega-\cos2\pi\omega)+(h_1+h_2)\,\omega\sin2\pi\omega={}\\
{}=h_1h_2\sin\omega\sigma_1\cdot\sin\omega(2\pi-\sigma_1).$$

Other relations connecting these parameters we obtain after
substituting expression (\ref{X}) into the orthonormality conditions
(\ref{ort}):
 \bea
 &\om^2(A_i^2+B_i^2+\tilde A_i^2+\tilde B_i^2)=
 a_0^2(1+\theta^2),
\qquad i=1,\dots,n;&\label{ABort1}\\
&\om^2(\tilde A_iB_i-A_i\tilde B_i)= a_0^2\theta,\qquad
i=1,\dots,n.&
 \label{ABort2}\eea

Among $n$ equations (\ref{ABort2}) only one is independent, for
example, with $i=1$. If it is satisfied and the relations
(\ref{AiBi}) take place, other conditions (\ref{ABort2}) is
satisfied too. But $n$ equations (\ref{ABort1}) are independent.
Below we use the first of them and their residuals
 \be
\breve C_i(h_i\breve C_i{+}2\breve S_i)(A_i^2{+}\tilde A_i^2)+
\breve S_i(h_i\breve S_i{-}2\breve C_i)(B_i^2{+}\tilde
B_i^2)=2(\breve C_i^2-\breve S_i^2-h_i\breve C_i\breve
S_i)(A_iB_i+\tilde A_i\tilde B_i).
 \label{ABort3}\ee
 Here Eqs.~(\ref{AiBi}) are used.

Under condition (\ref{eqC}) the matrix $M=0$ in Eq.~(\ref{M0}) and
an arbitrary nonzero column ${\cal A}_1$ or ${\cal B}_1$ is its
eigenvector. It is connected with the rotational symmetry of the
problem. So one can express, for example, the column ${\cal B}_1$
via ${\cal A}_1$ (the latter may be taken arbitrarily):
 \be
B_1=\frac{-C_*A_1+S_\theta\tilde A_1}{S_*},\quad \tilde
B_1=-\frac{S_\theta A_1+C_*\tilde A_1}{S_*}.
 \label{BA}\ee
 Here
$$\begin{array}{c}
 C_*=\breve C-C_\theta-h_1C_1(\breve SC_1-S_1\breve C-h_2S_2S_3)-h_2\breve C_2S_3,\\
S_*=\breve S-h_1S_1(\breve SC_1-S_1\breve C-h_2S_2S_3)-h_2\breve
S_2S_3, \rule{0mm}{5mm}\end{array}\qquad \mbox{for \ } n=3,$$
 $$
 C_*=\breve C-C_\theta-h_1C_1S_2,\quad S_*=\breve S-h_1S_1S_2,\qquad
\mbox{for \ } n=2. $$

Coefficients (\ref{BA}) must satisfy equations
(\ref{ABort1})\,--\,(\ref{ABort3}), resulting from the
orthonormality conditions (\ref{ort}). After substitution
expressions (\ref{BA}) into Eqs.~(\ref{ABort1}), (\ref{ABort2})
with $i=1$ we have
 \begin{eqnarray}
&\!\!\!\!\omega^2S_*^{-2}(C_*^2+S_*^2+S_\theta^2)(A_1^2+\tilde
A_1^2)=a_0^2(1+\theta^2),\;\;&
 \label{a01}\\
&\omega^2S_*^{-1}S_\theta(A_1^2+\tilde A_1^2)=a_0^2\theta.&
 \label{a02}\end{eqnarray}

 If we exclude factors $A_1^2+\tilde
A_1^2$ and $a_0^2$ from this system, we obtain
 \be
\frac{1+\theta^2}\theta=
\frac{C_*^2+S_*^2+S_\theta^2}{S_*S_\theta}.
 \label{omth}\ee
 This equation determines values of the parameters  $\omega$
and $\theta$. In the case $n=2$ equation (\ref{omth}) takes the
form
 \be
\frac{1+\theta^2}\theta=\frac{2\breve S+(h_1+h_2)\,\breve
C-h_1h_2C_1S_2}{S_\theta}.
 \label{omth2}\ee

To determine values $\sigma_1,\dots,\sigma_{n-1}$ one should add
$n-1$ equations (\ref{ABort3}) to the system (\ref{eqCg}),
(\ref{omth}) and take into account Eqs.~(\ref{BA}).

In the case $n=2$ one equation (\ref{ABort3}) is reduced to the
simple form
 $$
 \sin\big[2\omega(\pi-\sigma_1)\big]=0.
 $$
 It determines a set of acceptable values $\sigma_1$:
 \be
 \sigma_1=\pi+\frac{\pi k}{2\omega},\quad k\in Z,\quad |k|<2\omega.
 \label{s1}\ee

If values $\omega$, $\theta$, $\sigma_i$ satisfy equations
(\ref{eqCg}), (\ref{ABort3}), (\ref{omth}), the expression
(\ref{X}) satisfies the system (\ref{clos})\,--\,(\ref{qq0}) and
describes an uniform rotation of the closed string with massive
points (rotational state). The shape of this string is a section
$t={}t_0={}$const of the world surface (\ref{X}). This shape is
the closed curve, composed from segments of a hypocycloid if and
only if the equalities (\ref{eqCg}), (\ref{omth}) are fulfilled.
This result is similar to the behavior of rotational states for
the string baryon model ``triangle'' \cite{PRTr}.

Hypocycloid is the curve drawing by a point of a circle (with
radius $r$) rolling inside another fixed circle with larger radius
$R$. In the case of solutions (\ref{X}) uniformly rotating
hypocycloidal segments of the string are joined at non-zero angles
in the massive points. The relation of the mentioned radii is
$$
\frac rR=\frac{1-|\theta|}2.
$$
 For solutions (\ref{X}) $|\theta|<1$.

This hypocycloidal string rotates in the $e_1,\,e_2$ plane at the
angular velocity $\Omega=\om/a_0$, the massive points move at the
speeds $v_i$ along the circles with radii $v_i/\Omega$. These
values are connected by the following equations, resulting from
Eqs.~(\ref{Qi}):
 \be a_0
=\frac{m_1Q_1}{\gamma \sqrt{1-v_1^2}}=\dots=\frac{m_n Q_n}{\gamma
\sqrt{1-v_n^2}}.
 \label{a0v}\ee
 Speeds $v_i$ are determined by Eqs.~(\ref{Qi}) and in the case
$n=2$ are equal
 \be
v_1^2=\theta\frac{\breve S-h_2S_1S_2}{S_\theta},\qquad
v_2^2=\theta\frac{\breve S-h_1S_1S_2}{S_\theta}.
 \label{v12}\ee

The rotating string may also have cusps (return points) of the
hypocycloid moving at the speed of light.

Values $\om$ and $\theta$ are determined from the system
(\ref{eqCg}), (\ref{omth}).  Solution of the system
(\ref{eqCg}), (\ref{omth}) (pairs $\om$, $\theta$) form some
countable set. Each pair corresponds to solution (\ref{X})
describing uniform rotation of the closed string with certain
topological type.

The rotational states (\ref{X}) in the case $\theta\ne0$ we shall
name ``hypocycloidal states''.

In the case when the parameter in Eq.~(\ref{theta}) equals zero
($\theta=0$), solutions (\ref{X}) describe rotational motions of
$n$ times folded string. It has a form of rotating rectilinear
segment. These motions are divided into two classes: (a) ``linear
states'' with all masses $m_i$ moving at nonzero velocities $v_i$
at the ends of the rotating rectilinear folded string, and (b)
``central states'' with one massive point (or some of them) placed
at the rotational center (Fig.~2).

 There are many topologically different types of linear, central
and hypocycloidal states (\ref{X}). They may be classified with the
number of cusps and the type of intersections of the hypocycloid
following Ref.~\cite{PRTr}. Note that in the considered model
(\ref{S}) the string does not interact with itself in a point of
intersection.

These topological configurations of the rotational states may be
classified by investigation of the massless $m_i\to0$ or
ultrarelativistic $v_i\to1$ limit for fixed $\gamma$ and $a_0$.
Analysis of equations  (\ref{eqCg}), (\ref{omth})\,--\,(\ref{v12})
shows that in the limit $m_i\to0$ the values $Q_i$ tend to
infinity, values $2\om$ and $2\theta\om$ tend to following integer
numbers:
 \be
n_1=\Big|\lim\limits_{m_i\to0}2\om\Big|, \qquad
n_2=\lim\limits_{m_i\to0}2\theta\om.
 \label{n1n2}\ee

Because of the inequality $|\theta|<1$  and Eq.~(\ref{eqC}),
resulting in the equality $(-1)^{n_1}=(-1)^{n_2}$ only the
following values of $n_1$ and $n_2$ are admissible:
 \be n_1\ge2;\qquad n_2=n_1-2,\;\;
n_1-4,\;\dots\;-(n_1-2).
 \label{n12con}\ee

The number $n_1$ is the number of cusps of the rotating hypocycloid
(including massive points), the number $n_2$ describes the shape of
this curve.

For exhaustive classification of topological types of the states
(\ref{X}) we are to specify the following set
$(n_1,n_2,k_1,\dots,k_{n-1})$, including the numbers (\ref{n1n2})
and $n-1$ integer parameters $k_1,k_2,\dots,k_{n-1}$, which
numerate positions of massive points $m_1,m_2,\dots,m_{n-1}$.

In the case $n=2$ only one parameter $k_1\equiv k$ is required, we
can take the value $k$ in Eq.~(\ref{s1}) for this purpose.
Eq.~(\ref{s1}) restricts admissible values $k$:
 \be
 k=n_1-2,\,n_1-4,\,\dots\,2-n_1.
\label{kcon}\ee

In Fig.~2 examples of rotational states (\ref{X}) with different
types $(n_1,n_2,k)$ are presented.

\begin{figure}[ht]
\includegraphics[width=20 cm, trim=10mm 20mm 0mm 65mm]{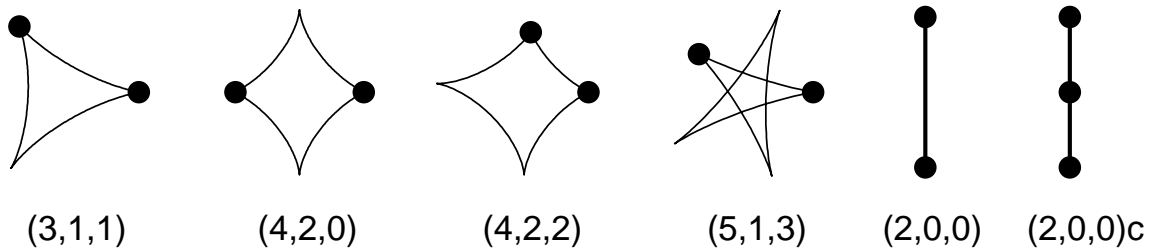}
\caption{Examples of rotational states of the type $(n_1,n_2,k)$}
\end{figure}

In the first 5 examples the string carries $n=2$ massive points,
in the last (the right) case the central state with $n=3$ is
shown.

The left state in Fig.~2 in the limit $m_i\to0$ tends to
hypocycloid with $n_1=3$ cusps, for the 2-nd and 3-rd states this
number is $n_1=4$ (they tend to astroid), but these curves have
different numbers $k$ in accordance with different positions of
masses. The fourth state with $n_1=5$, $n_2=1$ correspond to the
curvilinear star. The mentioned 4 examples present hypocycloid
states.

The fifth example with the type $(2,0,0)$ describes the simplest
linear rotational state with $n=2$. In the limit $m_i\to0$ this
state and the central state (the rightmost in Fig.~2) tend to the
same limit: the double rectilinear segment.

For linear and central rotational states we can put $\tilde A_1=0$
(without loss of generality) in the column ${\cal A}_1$. Equations
$S_\theta=0$,  (\ref{AiBi}), (\ref{ba0}), (\ref{BA}) result in
equalities $\tilde B_1=0$, $b_0=0$ and $\tilde u(\s)=0$. So the
parametrization (\ref{X}) of the string world surface for linear
and central rotational states may be rewritten in the simple form
 \be
X^\mu(\tau,\s)=x_0^\mu+e_0^\mu a_0\tau+ u(\s)\cdot e^\mu(\om\tau).
 \label{Xlin}\ee
 Parameters in Eq.~(\ref{Xlin}) are
determined from Eqs.~(\ref{AiBi})\,--\,(\ref{a0v}), but we are to
note that for the case $\theta=0$ equations (\ref{ABort2}) become
the identities and equations (\ref{omth}) or (\ref{omth2}) loose
their sense (and should be replaced by the equation $\theta=0$).
The values $\om$, $\s_i$ are determined from Eqs.~(\ref{eqCg}),
(\ref{ABort3}).

In the case $n=2$ for linear states (\ref{Xlin}) the value $\s_1$
is determined from Eq.~(\ref{s1}) as before (with arbitrary even
number $k$), and  equation (\ref{eqCg}) has the form (\ref{eqC}).
If we substitute in this  equation $C_\theta=1$ and equalities
$S_2=(-1)^k S_1=S_1$, $2S_1^2=1-\breve C$, resulting from
Eq.~(\ref{s1}), we obtain for linear states
 \be
(h_1h_2-4)\,\tan\pi\omega=2(h_1+h_2),
 \label{eqlin}\ee
Speeds $v_i$ of massive points for linear states are
 \be
v_i=2(4+h_i^2)^{-1/2}.
 \label{vi}\ee
 These relations result from Eqs.~(\ref{ABort1}), (\ref{BA}), (\ref{s1}), (\ref{eqlin})
 in the case $\theta=0$.

The central rotational states with $\theta=0$, $n_2=0$  are
described by Eq.~(\ref{Xlin}), but some massive points are placed
at the rotational center. For example, consider the central state
with $n=3$ massive points (the rightmost in Fig.~2), where the
mass $m_3$ is the center ($v_3=0$) and masses $m_1$, $m_2$ move at
nonzero speeds $v_1$ and $v_2$. The equality $v_3=0$ results in
the condition $u(0)=0$. It is equivalent to the equality
 \be
 A_1=0,
 \label{A1}\ee
 that forbids to use Eqs.~(\ref{BA}) for this state.

So we express coefficients $A_2$, $B_2$, $A_3$, $B_3$ via $B_1$
from Eqs.~(\ref{AiBi}), for example, $A_3=(h_1S_1^2+h_2\breve
S_2^2-h_1h_2\breve S_2S_1S_2)\,B_1$ and substitute these
expressions into Eqs.~(\ref{ABn}). Keeping in mind Eq.~(\ref{A1}),
we obtain two equations, connecting values $\om$, $\s_1$, $\s_2$,
$h_1$, $h_2$. These equations after transformations take the form
 \be
S_1+S_2C_3+C_2S_3=h_2S_2S_3,\qquad S_3+S_1C_2+C_1S_2=h_1S_1S_2.
 \label{eqs13}\ee
 Equation (\ref{eqCg}) is the consequence of Eqs.~(\ref{eqs13}).

Other relations between the mentioned values result from
Eqs.~(\ref{ABort3}) with $i=1$ and $i=2$. They may be reduced to
the form
 \be
h_1=2\frac{C_1}{S_1}=2\cot\om\s_1,\qquad
h_2=2\frac{C_3}{S_3}=2\cot\om(\pi-\s_1).
 \label{h1h2}\ee
 Eqs.~(\ref{eqs13}) and (\ref{h1h2}) result in the equality
 \be
\s_2-\s_1=\pi,
 \label{s21}\ee
and its consequences $C_2=C_1C_3-S_1S_3=\cos\pi\om$,
$S_2=S_1C_3+C_1S_3=\sin\pi\om$, $\breve S=2S_2C_2$.

Keeping in mind these relations we determine all coefficients of
the function $u(\s)$:
$$
A_2=2S_1C_1B_1,\quad B_2=(S_1^2-C_1^2)B_1,\quad A_3=-\breve
SB_1,\quad B_3=\breve CB_1,
$$
 the value $a_0=\om B_1$ from Eq.~(\ref{ABort1}), and, considering $\dot
X^\mu$ at $\s=\s_1$, $\s=\s_2$ and Eq.~(\ref{s21}), determine
velocities of the massive points:
 \be
v_1=S_1,\qquad v_2=S_3.
  \label{v12c}\ee

These equalities and Eqs.~(\ref{h1h2}) let us to express
$$h_i=2\sqrt{v_i^{-2}-1},\qquad i=1,2$$
 (coinciding with Eq.~(\ref{vi})) and, taking into account Eqs.~(\ref{a0v}),
 to obtain the equation
 \be
\frac{m_1v_1}{1-v_1^2}=\frac{m_2v_2}{1-v_2^2}.
 \label{mivi}\ee

If the initial data for this central rotational state are the
values $m_1$, $m_2$, $\gamma$, $v_1$, than one can find $v_2$ from
Eq.~(\ref{mivi}), and from Eqs.~(\ref{h1h2}) the values $h_1$,
$h_2$ and
 \be
\om=\frac1\pi\left(\arctan\frac2{h_1}+\arctan\frac2{h_2}\right)+n_1^*;\qquad
\s_1=\frac1\om\left(\arctan\frac2{h_1}+k\right),
 \label{oms1}\ee
and all other parameters of the world surface (\ref{Xlin}).

\bigskip

\centerline {\bf 4. Stability problem for central rotational
states}
\medskip

Possible applications of solutions (\ref {X}) and (\ref {Xlin}) in
hadron spectroscopy essentially depend on stability or instability
of these states with respect to small disturbances. In this
section we study spectrum of these disturbances for the central
rotational states.

This problem has been recently solved for the closed string with
$n=1$ massive point for central states in Ref.~\cite{clstab05},
and for linear and hypocycloidal states in Ref.~\cite{clhyp06}.
Here we generalize this approach to the case of larger numbers $n$
($n\le3$).

To solve the stability problem for rotational states (\ref{X}) or
(\ref{Xlin}) we consider the general solution of Eq.~(\ref{eq})
for the string with $n$ masses
 \be
X^\mu (\tau,\s)=\frac{1}{2}[\Psi^\mu_{i+}(\tau +\s)+\Psi
^\mu_{i-}(\tau-\s)],]\qquad\s\in[\s_{i-1},\s_i],\quad i=1,\dots,n.
 \label{gensol}\ee
 Here the functions $\Psi^\mu_{i\pm}(\tau\pm\s)$ are smooth, the world surface
(\ref{gensol}) is smooth between world lines of massive points.

We denote $\breve\Psi^{\mu}_{i\pm}$ the functions in the
expression (\ref {gensol}) for the rotational states (\ref{X}) or
(\ref{Xlin}). In particular, for  the central rotational state
(\ref{Xlin}) with $n=3$ massive points (the rightmost in Fig.~2),
where the mass $m_3$ is at the center, and equalities
(\ref{A1})\,--\,(\ref{oms1}) take place, the derivatives of
functions $\breve\Psi^{\mu}_{i\pm}$ are
 \be \begin{array}{l}
\breve\Psi^{\pr\mu}_{1\pm}(\tau)=a_0\Big[e_0^\mu\pm
e^\mu(\om\tau)\Big],\\
 \breve\Psi^{\pr\mu}_{2\pm}(\tau)=a_0\Big[e_0^\mu+2v_1C_1\acute
e^\mu(\om\tau)\pm(2v_1^2-1)\,e^\mu(\om\tau)\Big],\rule{0mm}{6.5mm}\\
 \breve\Psi^{\pr\mu}_{3\pm}(\tau)=a_0\Big[e_0^\mu-\breve S\acute
e^\mu(\om\tau)\pm\breve C e^\mu(\om\tau)\Big],\rule{0mm}{6.5mm}
\end{array}
 \label{Psic}\ee

To describe any small disturbances of the rotational motion, that
is motions close to states (\ref{X}) or (\ref{Xlin}) we consider
vector functions $\Psi^{\pr\mu}_{i\pm}$ close to
$\breve\Psi^{\pr\mu}_{i\pm}$ in the form
 \be
\Psi^{\pr\mu}_{i\pm}(\tau)=\breve\Psi^{\pr\mu}_{i\pm}(\tau)
+\f_{i\pm}^\mu(\tau).\label{Psi+f}\ee

The disturbance $\f_{i\pm}^\mu(\tau)$ is supposed to be small, so
we omit squares of $\f_{i\pm}$ when we substitute the expression
(\ref{Psi+f}) into dynamical equations (\ref{clos}), (\ref{qqi})
and (\ref{qq0}). In other words, we work in the first linear
vicinity of the states (\ref{X}) or (\ref{Xlin}). Both functions
$\Psi^{\pr\mu}_{i\pm}$ and $\breve\Psi^{\pr\mu}_{i\pm}$ in
expression (\ref{Psi+f}) must satisfy the condition
 $${\Psi'_{i+}\!\!}^2={\Psi'_{i-}\!\!}^2=0,$$
 resulting from Eq.~(\ref{ort}), hence in the first order approximation on $\f_{i\pm}$
the following scalar product equals zero:
 \be
\big(\breve\Psi^\pr_{i\pm},\f_{i\pm}\big)=0.
 \label{Psif}\ee

For the disturbed motions the equalities (\ref{sigmi})
$\s_i={}$const and (\ref{theta}) $\tau^*=\tau+2\pi\theta$,
generally speaking, is not carried out and should be replaced with
the equalities
 \be
\s_1(\tau)=s_1+\de_1(\tau),\qquad\s_2(\tau)=s_2+\de_2(\tau),
\qquad\tau^*=\tau+2\pi\theta+\de(\tau),
 \label{taudel}\ee
 where $\de_i(\tau)$ and $\de(\tau)$ are small disturbances. In
the case of the central states (\ref{Psic}) $\theta=0$.

Expression (\ref{Psi+f}) together with Eq.~(\ref{gensol}) is the
solution of the  string motion equation (\ref {eq}). Therefore we
can obtain equations of evolution for small disturbances
$\f_{i\pm}^\mu(\tau)$, substituting expressions (\ref{Psi+f}) and
(\ref{taudel}) with Eq.~(\ref{Psic}) into other equations of
motion (\ref{qqi}), (\ref{qq0}), the closure condition
(\ref{clos}) and the continuity condition
 \be
X^\mu\big(\tau,\s_i(\tau)-0\big)=X^\mu\big(\tau,\s_i(\tau)+0\big),\qquad
i=1,\dots,n-1.
 \label{cont}\ee
 We are to take into account nonlinear factors
$\Big\{\big[\frac d{d\tau}X(\tau,\s_i(\tau))\big]^2\Big\}^{-1/2}$
and contributions from the disturbed arguments $\tau^*$ and
$\s_i(\tau)$ (\ref{taudel}), for example:
$$
\breve\Psi^{\pr\mu}_{n\pm}(\tau^*\pm2\pi)\simeq
\breve\Psi^{\pr\mu}_{n\pm}(\tau+2\pi\theta\pm2\pi)+\de(\tau)\,
\breve\Psi^{\pr\pr\mu}_{n\pm}(\tau+2\pi\theta\pm2\pi).
$$

This substitution for the central rotational state (\ref{Xlin})
with $n=3$ and vector-functions $\breve\Psi^{\mu}_{i\pm}$
(\ref{Psic}) after simplifying results in the following system of
6 vector equations in linear (with respect to $\f_\pm^{i\mu}$,
$\de_i$ and $\de$) approximation:
 \be
\begin{array}{c}
\f_{1+}^\mu(+_1)+\f_{1-}^\mu(-_1)-\f_{2+}^\mu(+_1)-\f_{2-}^\mu(-_1)+4C_1a_0
\big[e^\mu(\om\tau)\,\dot\de_1(\tau)+\om\acute
e^\mu(\om\tau)\,\de_1\big]=0,\\
\f_{2+}^\mu(+_2)+\f_{2-}^\mu(-_2)-\f_{3+}^\mu(+_2)-\f_{3-}^\mu(-_2)-4C_3a_0
\big[e^\mu(\om\tau)\,\dot\de_2(\tau)+\om\acute
e^\mu(\om\tau)\,\de_2\big]=0,\rule{0mm}{5.5mm}\\
\f_{3+}^\mu(+)+\f_{3-}^\mu(-)-\f_{1+}^\mu(\tau)-\f_{1-}^\mu(\tau)+2a_0e_0^\mu
\dot\de(\tau)=0,\rule{0mm}{5.5mm}\\
\frac d{d\tau}\Big\{\f_{1+}^\mu(+_1)+\f_{1-}^\mu(-_1)+2C_1a_0
(e^\mu\dot\de_1+\om\acute e^\mu\de_1)+F_1(e_0^\mu+v_1\acute
e^\mu)\Big\}+{}
\rule{0mm}{6.8mm}\\
\qquad{}+Q_1\Big[\f_{1+}^\mu(+_1)-\f_{1-}^\mu(-_1)-\f_{2+}^\mu(+_1)+\f_{2-}^\mu(-_1)\Big]=0.
\rule{0mm}{6.0mm}\\
\frac d{d\tau}\Big\{\f_{2+}^\mu(+_2)+\f_{2-}^\mu(-_2)-2C_3a_0
(e^\mu\dot\de_2+\om\acute e^\mu\de_2)+F_2(e_0^\mu-v_2\acute
e^\mu)\Big\}+{}
\rule{0mm}{6.80mm}\\
\qquad{}+Q_2\Big[\f_{2+}^\mu(+_2)-\f_{2-}^\mu(-_2)-\f_{3+}^\mu(+_2)+\f_{3-}^\mu(-_2)\Big]=0.
\rule{0mm}{6.0mm}\\
\!\!\!\frac
d{d\tau}\Big\{\f_{1+}^\mu+\f_{1-}^\mu+(\f_{1+}-\f_{1-})\,e_0^\mu\Big\}+
Q_3\Big[\f_{3+}^\mu(+)-\f_{3-}^\mu(-)-\f_{1+}^\mu+\f_{1-}^\mu+2\om
a_0\acute e^\mu\de\Big]=0. \rule{0mm}{6.5mm}\!\!\!
\end{array} \label{sysf}\ee
 Here arguments $(\tau)$ for $\f_{1\pm}^\mu$, $\de$, $\de_i$ and $(\om\tau)$ for $e^\mu$,
$\acute e^\mu$ may be omitted; we use the following notations for
arguments
 $$
(\pm_1)\equiv(\tau\pm\s_1),\qquad
(\pm_2)\equiv(\tau\pm\s_2),\qquad(\pm)\equiv(\tau\pm2\pi),
 $$
 for the scalar products
 \be
  \f_{i\pm}^0\equiv( e_0,\f_{i\pm}),\qquad
\f_{i\pm}^3\equiv( e_3,\f_{i\pm}),\qquad \f_{i\pm}\equiv (
e,\f_{i\pm}), \qquad \acute\f_{i\pm}\equiv(\acute e,\f_{i\pm})
\label{fiscal}\ee
 and
 $$
\begin{array}{l}
F_1=\f_{1+}(+_1)-\f_{1-}(-_1)-v_1C_1^{-1}\Big[\acute\f_{1+}(+_1)+\acute\f_{1-}(-_1)-
2\om a_0\de_1\Big],\\
F_2=C_3^{-1}\Big\{\breve C_2\big[\f_{2-}(-_2)-\f_{2+}(+_2)\big]+
\breve S_2\big[\acute\f_{2+}(+_2)+\acute\f_{2-}(-_2)\big]+ 2\om
v_2a_0\de_2\Big\}.\rule{0mm}{6.0mm}
\end{array}
 $$

The first two equations (\ref{sysf}) results from
Eqs.~(\ref{cont}), the third ---  from Eq.~(\ref{clos}), other
ones are consequence of Eqs.~(\ref{qqi}) and (\ref{qq0}).
Equations (\ref{sysf}) are simplified with using
Eqs.~(\ref{A1})\,--\,(\ref{oms1}), (\ref{Psic}) and equalities
(\ref{Psif}), resulting in the following relations for projections
(\ref{fiscal}) of disturbances:
 \be
\f_{1\pm}^0(\tau)\pm\f_{1\pm}(\tau)=0,\quad
\f_{2\pm}^0+2v_1C_1\acute\f_{2\pm}\pm(2v_1^2-1)\,\f_{2\pm}=0,\quad
\f_{3\pm}^0-\breve S\acute\f_{3\pm}\pm\breve C\f_{3\pm}=0.
 \label{eqscal}\ee

The linearized system of equations (\ref{sysf}), (\ref{eqscal})
describes evolution of small disturbances of the considered
central rotational state (\ref{Xlin}), (\ref{Psic}).

Note that scalar products of Eqs.~(\ref{sysf}) onto the vector
$e_3$ (orthogonal to the rotational plane $e_1$,\,$e_2$) form the
closed subsystem from 6 equations with respect to 6 functions
(\ref{fiscal}) $\f_{i\pm}^3$:
 \be
\begin {array}{c}
\f_{i+}^3(+_i)+\f_{i-}^3(-_i)=\f_{i^*+}^3(+_i)+\f_{i^*-}^3(-_i),\\
\f_{3+}^3(+)+\f_{3-}^3(-)=\f_{1+}^3(\tau)+\f_{1-}^3(\tau),\\
\dot\f_{i+}^3(+_i)+\dot\f_{i-}^3(-_i)+Q_i\Big[
\f_{i+}^3(+_i)-\f_{i-}^3(-_i)-\f_{i^*+}^3(+_i)+\f_{i^*-}^3(-_i)\Big]=0,
\rule{0mm}{5mm}\\
 \dot\f_{1+}^3(\tau)+\dot\f_{1-}^3(\tau)+Q_3\Big[
\f_{3+}^3(+)-\f_{3-}^3(-)-\f_{1+}^3(\tau)+\f_{1-}^3(\tau)\Big]=0.
\rule{0mm}{5mm}\end{array}
 \label{sysf3}\ee
 Here $i=1,2$, $i^*\equiv i+1$. This system is homogeneous system
with deviating arguments.

We search solutions of this system in the form of harmonics
 \be
\f_{j\pm^3}=B_{j\pm}^3 e^{-i\tom\tau}.
 \label{fexp3} \ee

This substitution results in the linear homogeneous system of 6
algebraic equations with respect to 6 amplitudes $B_{i\pm}^3$. The
system has nontrivial solutions if and only if its determinant
 $$ \left|
\begin{array}{cccccc}
E_{1+} & E_{1-}& -E_{1+} & -E_{1-}& 0 & 0 \\
 0 & 0 & E_{2+} & E_{2-}& -E_{2+} & -E_{2-}\\
-1 & -1 & 0 & 0 & E_{3+} & E_{3-}\\
(i\tom-Q_1)\,E_{1+}& (i\tom+Q_1)\,E_{1-}&Q_1E_{1+}& -Q_1E_{1-} & 0 & 0 \\
 0 & 0 & (i\tom-Q_2)\,E_{2+}& (i\tom+Q_2)\,E_{2-}& Q_2E_{2+} &-Q_2E_{2-} \\
-i\tom-Q_3 &-i\tom+Q_3 & 0 & 0 & Q_3E_{3+} &-Q_3E_{3-}
\end{array}\right|=0
 $$
 equals zero. Here $E_{j\pm}=\exp(\mp i\tom\s_j)$. This equation is reduced to
 the form
 $$
2(1-\cos2\pi\tom)+\tom\big(Q_1^{-1}+Q_2^{-1}+Q_3^{-1}\big)\sin2\pi\tom={}
 $$
 \be
{}=\tom^2\left(\frac{\sin^2\pi\tom}{Q_1Q_2}+
\frac{\sin\tilde\s_3\tom\cdot\sin\s_2\tom}{Q_2Q_3}+
\frac{\sin\s_1\tom\cdot\sin\tilde\s_{23}\tom}{Q_1Q_3}\right)-
\frac{\tom^3\sin\s_1\tom\cdot\sin\pi\tom\cdot\sin\tilde\s_3\tom}{Q_1Q_2Q_3},
 \label{tom3}\ee
 where $\tilde\s_3=2\pi-\s_2=\pi-\s_1$, $\tilde\s_{23}=2\pi-\s_1$.
This equation coincides with Eq.~(\ref{eqCg}) with $n=3$, if $\om$
is substituted by $\tom$. The spectrum of transversal (with
respect to the $e_1,\,e_2$ plane) small fluctuations of the string
for the considered rotational state contains frequencies $\tom$
which are roots of Eq.~(\ref{tom3}). Analysis of the real and
imaginary parts of this equation demonstrates that all these
frequencies are real numbers, therefore amplitudes of such
fluctuations do not grow with growth of time $t$.

Let us consider small disturbances concerning to the $e_1,\,e_2$
plane. Projections (scalar products) of equations (\ref{sysf})
onto 3 vectors $e_0$, $e(\tau)$, $\acute e(\tau)$ form the system
of 18 differential equations with deviating arguments with respect
to 15 unknown functions of the argument $\tau$: $\f_{j\pm}$,
$\acute\f_{j\pm}$ ($j=1,2,3$), $\de_1$, $\de_2$, $\de$ (functions
$\f_{j\pm}^0$ are excluded via Eqs.~(\ref{eqscal})).

When we search solutions of this system in the form of harmonics
(\ref{fexp3})
 \be
\f_{j\pm}^0=B_{j\pm}^0 e^{-i\tom\tau},\qquad\f_{j\pm}=B_{j\pm}
e^{-i\tom\tau},\qquad \acute \f_{j\pm}=\acute B_{j\pm}
e^{-i\tom\tau}, \qquad 2a_0\de_j=\Delta_j e^{-i\tom\tau},
 \label{fexp} \ee
 we obtain the homogeneous system of 18 algebraic equations with
respect to 15 amplitudes $B_{j\pm}^0$, $B_{j\pm}$, $\acute
B_{j\pm}^0$, $\Delta_1$, $\Delta_2$, $\Delta$. Three of these 18
equations are linear combinations of other ones.

For the rest 15 equations we use the mentioned above condition of
existence of nontrivial solutions for this system. It is vanishing
the corresponding determinant. These equations and calculations
are cumbersome, so we omit this system and present here the result
of symbolic calculation in the package MATLAB. In the case
$m_1=m_2$, $\s_1=\tilde\s_3=\pi/2$ the condition of vanishing this
determinant is reduced to the following equation:
 \be
 4Q_3^2\tan^2\pi\tom
+4Q_3\left(\tom+\frac{\om^2}{2\tom}\right)\tan\pi\tom+\tom^2-\om^2=0.
 \label{eqtom} \ee
  This equation generalizes the condition in Ref.~\cite{clstab05}
for the closed string with $n=1$ massive point. It transforms
 into the mentioned condition in the limit $m_1\to0$, $m_2\to0$.
  The transcendental equation (\ref{eqtom}) contains a
denumerable set of real roots (frequencies). They correspond to
different modes of small oscillations of the string in the
considered central rotational state (\ref{Xlin}).

This state will be unstable, if there are complex frequencies
$\tom=\check\om+i\xi$ in the spectrum, generated by
Eq.~(\ref{eqtom}). If its imaginary part $\xi$ will be positive,
the modes of disturbances $\f^\mu$ (corresponding to the root
$\check\om+i\xi$) get the multiplier $\exp(\xi\tau)$, that is they
grow exponentially.

The search of complex roots of equation (\ref{eqtom}) in
Ref.~\cite{clstab05} showed that such roots can exist only on the
imaginary axis of the complex plane $\tom$. On this axis of $\tom$
(in the case $\tom=i\xi$) the equation (\ref{eqtom}) takes
the form
 \be
4Q_3^2\tanh^2\pi\xi
+\xi^2+\om^2=4Q_3\left(\frac{\om^2}{2\xi}-\xi\right) \tanh\pi\xi.
 \label{eqxi}\ee
  The left hand side of this equation grows with
growing $\xi$ (for $\xi>0$), and the right hand side decreases. It
is obvious (see the limit $\xi\to0$), that the root $\xi>0$ of
Eq.~(\ref{eqxi}), that is the imaginary root $\tom=i\xi$ of
Eq.~(\ref{eqtom}) exists, if and only if
 \be
 2\pi Q_3>1.
  \label{crit1} \ee

If we use the expression (\ref{a0v}) in the form $Q_3=\gamma
a_0/m_3$ (remind that $v_3=0$ for this central state) we
reduce the criterion (\ref{crit1}) to the following form:
 \be
m_3<m_{cr}\equiv 2\pi\gamma a_0.
 \label{crit2}\ee

Thus, we obtain the threshold effect in stability properties of
the central rotational states under consideration. If the central
mass $m_3$ is greater than $m_{cr}$, hence all roots of
Eq.~(\ref{eqtom}) are real ones and the state is stable. But in
the case $m_3<m_{cr}$ the state is unstable: the imaginary root
$\tom=i\xi$ appears and the corresponding amplitude of
disturbances grows exponentially: $\f=B e^{\xi\tau}$.

This threshold effect or the spontaneous symmetry breaking for the
string state was observed in numerical experiments in
Ref.~\cite{clstab05}. Note that our analysis of small disturbances
is suitable only for initial stage of an unstable motion when
disturbances are really small.

In the following section the stable rotational states are applied
in hadron spectroscopy.

\bigskip

\centerline {\bf 5. Regge trajectories}
\medskip

The obtained rotational motions of the considered model should be
applied for describing physical manifestations of glueballs and
other exotic particles, in particular, their Regge trajectories.
For this purpose we calculate the energy $E$ and classic angular
momentum $L$ for the states (\ref{X}) of this model. For an
arbitrary classic state of the relativistic string with the action
(\ref{S}) carrying  massive points they are determined by the
following integrals (Noether currents) \cite{PRTr,MilSh}:
 \bea &\displaystyle P^\mu=\int\limits_{\cal C}
p^\mu(\tau,\s)\,d\s +\sum_{i=1}^np_i^\mu(\tau),&
 \label{Pimp}\\
&\displaystyle {\cal L}^{\mu\nu}=\int\limits_{\cal
C}\Big[X^\mu(\tau,\s)\,
p^\nu(\tau,\s)-X^\nu(\tau,\s)\,p^\mu(\tau,\s)\Big]\,d\s+\sum_{i=1}^n
\Big[x_i^\mu(\tau)\,p_i^\nu(\tau)-x_i^\nu(\tau)\,p_i^\mu(\tau)\Big],&
 \label{Mom}\eea
 where $x_i^\mu(\tau)=X^\mu\big(\tau,\s_i(\tau)\big)$ and
$p_i^\mu(\tau)=m_i\dot x_i^\mu(\tau)\big/\sqrt{\dot x_i^2(\tau)}$
are coordinates and momentum of the massive points,
 $p^\mu(\tau,\s)=\gamma\big[(\dot X,X') X^{\pr\mu}-X'{}^2\dot
X^\mu\big]/\sqrt{-g}$
 is the canonical string  momentum,
 ${\cal C}$ is
any closed curve (contour) on the tube-like world surface of the
string. Note that the lines $\tau={}$const on the world surface
(\ref{X}) are not closed in the case $\tau_0\ne0$. So we can use
the most suitable lines $\tau-\theta\s={}$const (that is
$t={}$const) as the contour $C$ in integrals (\ref{Pimp}),
(\ref{Mom}).

The reparametrization $\tilde\tau=\tau-\theta\s$, $\tilde\s=\s
-\theta\tau$ keeps the orthonormality conditions (\ref{ort}).
Under them $p^\mu(\tau,\s)=\gamma\dot X^\mu(\tau,\s)$.

The square of energy $E^2$ equals the scalar square of the
conserved vector of momentum (\ref{Pimp}): $P^2=P_\mu P^\mu=E^2$.
If we substitute the expressions (\ref{X}), (\ref{AiBi}),
(\ref{ba0}), (\ref{BA})\,--\,(\ref{a02})  into Eq.~(\ref{Pimp}) we
obtain the following formula for the momentum:
 \be
P^\mu=e_0^\mu E,\qquad E=2\pi\gamma  a_0(1-\theta^2)+\sum_{i=1}^n
\frac{m_i}{\sqrt{1-v_i^2}}.
 \label{P}\ee

For the classical angular momentum (\ref{Mom}) only $z$-component
of ${\cal L}^{\mu\nu}$ is nonzero:
 \be
{\cal L}^{\mu\nu}= \ell_3^{\mu\nu}L,\qquad
 L=\frac{\gamma
a_0^2}{2\omega}\left[2\pi(1-\theta^2)+ \sum_{i=1}^n
\frac{v_i^2}{Q_i} \right].
 \label{L}\ \ee
 Here $j_3^{\mu\nu}=e_1^\mu e_2^\nu-e_1^\nu
e_2^\mu=e^\mu\acute e^\nu- e^\nu\acute e^\mu.$

One can obtain the total angular momentum $J=L+S$ from the
classical momentum (\ref{L}) after quantization the system. But
this problem for the string with masses (\ref{S}) is not solved
yet because of essential nonlinearity of equations (\ref{qqi}),
(\ref{qq0}). So we use below the approach, suggested in
Refs.~\cite{Ko,4B} for string models of mesons and baryons. It
includes the spin contribution to the classical angular momentum
in the form
 \be
 J=L+S,\qquad S=\sum_{i=1}^ns_i,
 \label{J}\ee
 where $s_i$ are spin projections of massive points (valent glueballs),
and also the following contribution to the energy (\ref{P})
because spin-orbit interaction \cite{4B}:
 \be
\Delta E_{SL}=\sum_{i=1}^n\big[1-(1-v_i^2)^{1/2}\big] ({\boldmath
\Omega}\cdot{\boldmath s}_i).
 \label{corrm}\ee

Below we suppose that the value $S$ in Eq.~(\ref{J}) corresponds
to the maximal total momentum (\ref{J}), that is $S=2$ for 2-gluon
glueballs \cite{AbreuB}. Other values of model parameters are:
 \be
\gamma=0{.}175\mbox{ CeV}^2,\quad m_1=m_2=700\mbox{ MeV}.
 \label{gamm12}\ee
 This tension $\gamma$  corresponds to the slope of Regge
trajectories (\ref{Reggm}) for hadrons $\alpha'\simeq0{.}9$
GeV$^{-2}$. Estimations of gluon masses on the base of gluon
propagator  \cite{Bernard}, in particular, in lattice calculations
\cite{BonnBLW,SilvaO} yield values $m_i$ from 700 to 1000 MeV.

If the values $m_i$, $\gamma$ and the topological type
$(n_1,n_2,k_j)$ of the rotational state (\ref{X}) are fixed we
obtain the one-parameter set of motions with different values $E$
and $J$. These states lay at quasilinear Regge trajectories. As
the parameter of this set one can use any of these values:
$\omega$, $\theta$,  $a_0$, $E$, $J$ et al. Other values may be
expressed from Eqs.~(\ref{eqC})\,--\,(\ref{v12}).

In particular, in the case $n=2$ and $m_1=m_2$ (important for
applications for glueballs) the equalities $h_1=h_2$, $v_1=v_2$
take place, and the system (\ref{eqC}), (\ref{omth2}) is reduced
to one equation
 \be
S_1^2\Sigma_\theta^2+2(S_2+S_1C_\theta)
(C_1\Sigma_\theta-S_2)=C_1^2S_\theta^2,
 \label{eqCom}\ee
 where $\Sigma_\theta=\frac12S_\theta(1+\theta^2)/\theta$.

For every given value $\omega$ and fixed $k$ (this lets us to
obtain $\sigma_1$ from Eq.~(\ref{s1}) and also $S_i$, $C_i$) we
find $\theta$ from equation (\ref{eqCom}), then the values $h_i$,
$a_0$, $v_i$ from Eqs.~(\ref{eqC}), (\ref{omth2}), (\ref{a0v}),
(\ref{v12}). Using Eqs.~(\ref{P})\,--\,(\ref{corrm}) we obtain the
dependence $J=J(E^2)$ (the Regge trajectory).

Regge trajectories, calculated for rotational states (\ref{X}),
(\ref{Xlin}) of closed string with 2 massive points with different
topological types (see Fig.~2) are shown in Fig.~3 with the
corresponding type $(n_1,n_2,k)$. For all curves the values of
parameters $S=2$ and (\ref{gamm12}) are chosen, $J$ is in units
$\hbar$. The pomeron trajectory (\ref{ReggPom}) is shown as the
dashed line.

These Regge trajectories are nonlinear for small $E$ and tend to
linear if $E\to\infty$. Their slope in this limit depends on the
fixed topological type.

The ultrarelativistic limit $E\to\infty$ corresponds to
$v_i\to1-0$ (except for central states) and for values $\om$ and
$\theta$ --- to the limits (\ref{n1n2}). Substituting into
Eqs.~(\ref{eqC}), (\ref{omth}), (\ref{v12}),  (\ref{P}), (\ref{J})
asymptotic relations with small values $\ep_1=\sqrt{1-v_1^2}$,
$\ep_2=\sqrt{1-v_2^2}$, $2\om=n_1-\ep_\om$,
$n_1\theta=n_2-\ep_\theta$, we obtain in the limit $J\to\infty$,
$E\to\infty$ the following asymptotic relation between these
values for fixed type ($n_1,\,n_2,\,b_k$) of the state:
 \be
J\simeq\al'E^2+\al_1E^{1/2}+\alpha_2E^{-1/2}, \qquad E\to\infty,
 \label{JElim}\ee
 where
 \be
 \al'=\frac1{2\pi\gamma}\,\frac{n_1}{n_1^2-n_2^2},
 \label{als}\ee

\begin{figure}[ht]
\includegraphics[scale=1.15,trim=-20 10 10 10]{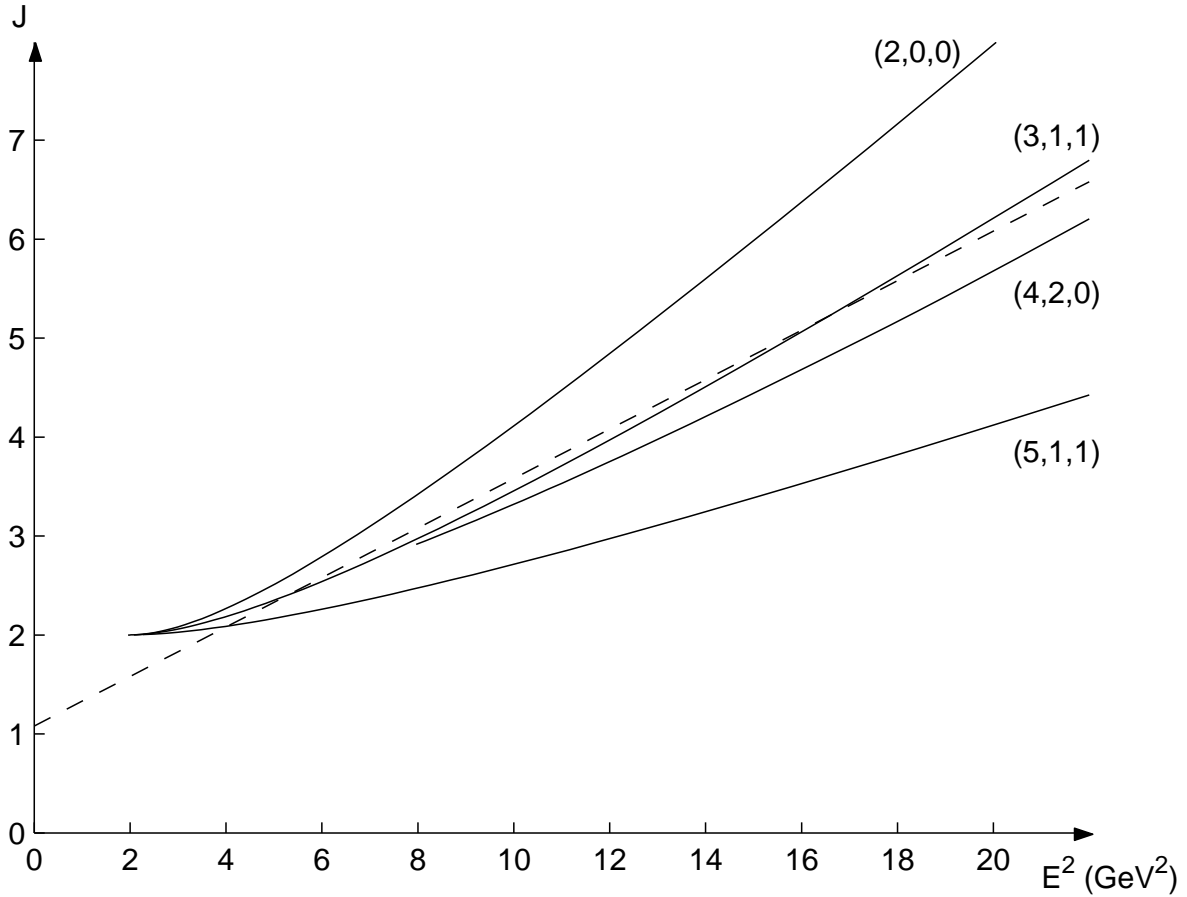}
\caption{Regge trajectories for rotational states (\ref{X}) with
different topological types}
\end{figure}

$$\al_1=-\frac{\sqrt2\,n_1(\sum m_i^{3/2})}{3\sqrt\pi\gamma(n_1^2-n_2^2)^{3/4}},\qquad
\alpha_2=\sqrt{\frac\pi2}(n_1^2-n_2^2)^{1/4} \sum_{i=1}^ns_i
m_i^{3/2}.$$

This dependence is close to linear one (\ref{Reggm}), but the
slope $\al'$  (\ref{als}) for this system differs from Nambu value
$\al'=1/(2\pi\gamma)$ by the factor
 \be
\chi=\frac{n_1}{n_1^2-n_2^2}.
 \label{chi}\ee
 In particular, the maximal slope with the factor
$\chi=1/2$ corresponds to the linear state with the type $(2,0,0)$
(two masses connected two strings without singularities). This
state's trajectory has the slope  (\ref{als})
$\alpha'\simeq0{.}45$ GeV$^{-2}$. It is larger that the slope
$\alpha'\simeq0{.}25$ of the pomeron trajectory (\ref{ReggPom})
(the dashed line in Fig.~3). These  trajectories distinctly
diverge in Fig.~3.

The Regge trajectories for central states are not shown in Fig.~3
because of their instability, studied in Sect.~4. This instability
takes place in the case (\ref{crit2}) that corresponds to
$E>3m_i$.

For chosen values (\ref{gamm12}) of gluon masses the most close to
the pomeron (glueball) trajectory (\ref{ReggPom}) is the
trajectory for the ``triangle'' configuration $(3,1,1)$. For this
state $\chi=3/8$ and the slope (\ref{als}) $\alpha'\simeq0{.}337$
GeV$^{-2}$. It is a bit larger than the value (\ref{ReggPom}), so
at very high energies $E$ these trajectories diverge. Some other
types of rotational states also generate suitable Regge
trajectories, for example, the state with $n_1=4$, $n_2=2$ gives
$\alpha'\simeq0{.}3$.

\bigskip

\centerline {\bf Conclusion}
\medskip

The obtained rotational states (\ref{X}) of the  closed string
with $n$ massive points are divided in 3 groups: hypocycloidal,
linear and central states, and also in a set of different
topological classes, described by the integer parameters
(\ref{n1n2}) $(n_1,n_2,k_1,\dots,k_{n-1})$. The states from these
classes generate the wide spectrum of quasilinear Regge
trajectories (\ref{P}), (\ref{J}) with different slopes
(\ref{als}) in the limit of large energies. This slope $\al'$
(\ref{als}) depends on only numbers (\ref{n1n2}) $n_1,n_2$,
describing the limiting shape of the closed string in the limit
$m_i\to0$ and does not depend on mutual positions of massive
points (numbers $k_j$).

For the central rotational states with the mass $m_3$ at the
rotational center and moving masses $m_1=m_2$ the stability with
respect to small disturbances is investigated. It is shown that
these states are unstable, if the central mass is less than the
critical value (\ref{crit2}). In this case the spectrum
disturbances has exponentially growing modes.

Regge trajectories (\ref{P}), (\ref{L}) for rotational states
(\ref{X}) states were calculated with spin corrections in the form
(\ref{J}), (\ref{corrm}).  There are some classes of hypocycloidal
rotational states (\ref{X}) suitable for describing the pomeron
(glueball) trajectory (\ref{ReggPom}), in particular the state
with $n_1=3$, $n_2=1$.

The considered model needs further development, in particular,
quantization or quantum corrections. These corrections are to be
significant for calculation of the intercept $\al_0$.

\medskip

\centerline{\bf Acknowledgment}

\medskip

The author is grateful to Russian foundation of basic research for
grant 05-02-16722.

\end{document}